\documentclass[%
superscriptaddress,
twocolumn,
nofootinbib,
amsmath,amssymb,
pre,
]{revtex4-1}

\usepackage{graphicx}
\usepackage{dcolumn}
\usepackage{bm}
\usepackage{hyperref}
\usepackage{amsmath} 
\usepackage{color}
\usepackage[dvipsnames]{xcolor}
\hypersetup{
	colorlinks,
	linkcolor={red!95!blue},
	citecolor={Cerulean!100},
	urlcolor={blue!80!black}
} 

\begin{document}
	
	\preprint{Submitted to PRE}
	
	\title{Feasibility study for implementing an optical Thomson scattering system for studying photoionized plasmas on Z}
	
	
	\author{P. M. Kozlowski}
	\email[]{pawel@pmkozlowski.com}
	\affiliation{Department of Physics, West Virginia University, Morgantown, WV 26506, USA}
	
	\author{R. C. Mancini}
	\affiliation{Department of Physics, University of Nevada, Reno, NV 89557-0058 }
	
	\author{M. E. Koepke}
	\affiliation{Department of Physics, West Virginia University, Morgantown, WV 26506, USA}
	
	\date{\today}
	
	\begin{abstract}
		Many astrophysical environments such as X-ray binaries, active galactic nuclei, and accretion disks of compact objects have photoionized plasmas. The strong photoionizing environment found near these bright X-ray sources can be produced in a scaled laboratory experiment, and direct measurements can form a testbed for spectroscopic models and photoionization codes used in analysis of these astrophysical objects. Such scaled experiments are currently being conducted using Ne filled gas cells on the Z-facility as part of the Z Astrophysical Plasma Properties (ZAPP) collaboration. The plasma is diagnosed using a pressure sensor for density and X-ray absorption spectroscopy for charge-state distribution. The electron temperature is presently inferred from a Li-like ion level population ratio, but it is necessary to obtain an independent temperature measurement, as photoionization alters the charge state distribution and can therefore cause errors in temperatures obtained via line ratio techniques. Optical Thomson scattering is a fitting diagnostic because it directly probes the distribution of plasma particle velocities with respect to a central probe frequency. It is a powerful diagnostic which can produce time and space resolved measurements of electron temperature, as well as, electron density, ion temperature, and average ionization. In this paper, we explore a possible design for an optical Thomson scattering system to supplement X-ray spectroscopic measurements. The proposed design will use equipment that is available on Z, though not yet assembled. Both the feasibility and impact of this new diagnostic are assessed by simulating expected spectra for a range of plasma parameters, thereby demonstrating the sensitivity of this diagnostic.
		
	\end{abstract}
	
	
	\maketitle
	
	\section{Introduction}
	
	Optical Thomson scattering (OTS) is a diagnostic where an optical probe beam interacts with a plasma, scattering from ion and electron density fluctuations within the plasma. The scattered spectrum is sensitive to a wide range of plasma parameters, such that the diagnostic has become a standard technique for measuring spatially localized ion/electron temperatures and densities, average ionization, and bulk plasma velocity, in both laser-produced and pulsed-power (pinch) plasmas  \cite{harvey2012optical,swadling2014diagnosing}. 
	
	To measure the electron plasma wave (EPW) and ion-acoustic wave (IAW) features simultaneously, the collected Thomson scattered light is divided using a beamsplitter and sent to two spectrometers, one with substantially higher resolution than the other in order to resolve the small shift in the ion feature \cite{ross2011ultraviolet, pollock2012simultaneous}. To measure the electron plasma wave feature, a notch filter is typically used to cut out the high signal from the ion feature such that it does not limit the resolution of the electron plasma wave feature due to the limited dynamic range of the spectrometer.
	
	This paper is laid out as follows. First, we describe the experiment of interest for which we wish to implement a Thomson scattering measurement. Second, an overview of Thomson scattering theory is given along with initial simulations of the expected spectral features. Third, the diagnostic design is discussed, including possible laser parameters, beamlines, collection optics, and spectrometers. Finally, all of the material from the prior sections is brought together to generate expected spectra (in terms of radiated power) for the given experiment, probe laser, and spectrometer arrangement.
	
	\section{Neon gas cell experiment}

	\begin{figure*}
		\centering
		\includegraphics[width=0.85\linewidth]{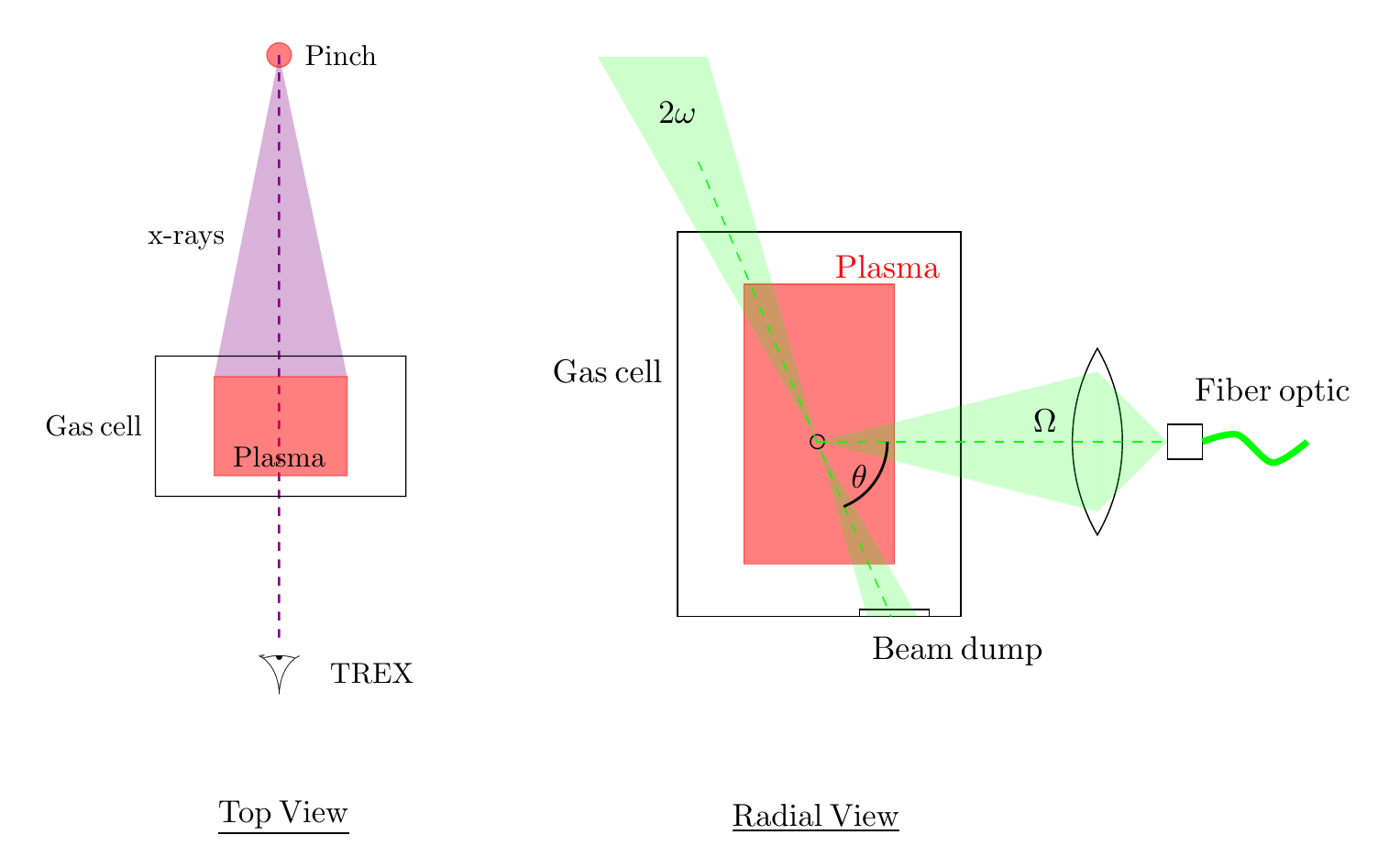}
		\caption{\emph{Left:} Top view of the photoionized Ne gas cell experiment. The gas cell is simultaneously heated and photoionized by the broad band X-ray flux from the pinch. The pinch also acts as a backlighter for the time-resolved x-ray spectroscopy (TREX) diagnostic. \emph{Right:} Radial view of the gas cell, where the pinch is located into the page. A possible Thomson scattering geometry is illustrated. The probe beam would enter via a window at the top of the gas cell and a collection optic would then collect the scattered light at a set angle $\theta$ over the solid angle $\Omega$.  The collection optic is coupled to a fiber optic cable, which then transports the Thomson scattered light to a spectrometer. The collection optic would be located in a separate cell (not illustrated), attached to the gas cell with a window between the two cells to protect it from the pinch radiation and gas cell plasma.}
		\label{fig:gascellschematic}
	\end{figure*}
	
	We will be examining the applicability of Thomson scattering to experiments on photoionized Neon conducted at the Z Facility at Sandia National Laboratories as part of the Z Astrophysical Plasma Properties (ZAPP) collaboration \cite{hall2014absorption}. The experiment consists of a gas cell filled with Neon radially offset from a Z pinch dynamic hohlraum (ZPDH), which acts as a broadband x-ray source \cite{rochau2014zapp}. The gas cell has a 1.4 micron thick Mylar window which allows these broadband x-rays to enter the gas cell and photoionize the Neon. The gas cell has been fielded at filling pressures of $P = 15 \:\rm Torr$ and $P = 30 \:\rm Torr$, corresponding to atomic densities of $n_a = 5 \times 10^{17} \:\rm cm^{-3}$ and $n_a = 1 \times 10^{18} \:\rm cm^{-3}$, respectively. A time-gated x-ray spectrometer is used to observe the transmission of a narrow band portion of the x-ray flux to record K-shell line absorption from Neon ions. The Neon ions areal densities extracted from the transmission spectrum permit the determination of the charge state distribution of the photoionized Neon plasma. From the ratio of level populations in ground and low-excited states of the Li-like ion the electron temperature of the plasma can be obtained \cite{mancini2018}. Initial results suggest an electron temperature of $T_e = 30 \:\rm eV$ and an average ionization $Z \approx 8$. However, it is important to also measure the electron temperature independent of x-ray spectroscopy.  It is therefore the motivation of this paper to examine the feasibility of supplementing X-ray spectroscopic measurements with optical Thomson scattering.
		
	Optical Thomson scattering provides a number of benefits over other diagnostics, including: simultaneous measurement of multiple plasmas parameters (electron and ion temperatures and densities), measurements localized to the scattering volume as opposed to line integrated measurements, and spectral features which are sensitive to the distribution function, and therefore the temperature of a plasma \cite{froula2006thomson}. This last point is important, as Thomson scattering does not require us to make an assumption about the distribution of the charge states based on the Saha-Boltzmann relation when inferring the temperature, but rather the temperature sensitivity derives from effects such as the Doppler broadening of Thomson scattering spectral features.

	\section{Thomson scattering theory}

	For a nonrelativistic plasma, the Thomson scattering spectrum for a given incident probe is described by the dynamic structure factor \cite{sheffield2010plasma}:
	\begin{equation}
		S \left(\vec{k}, \omega\right) = \frac{2 \pi}{k} \left|1 - \frac{\chi_e}{\epsilon}\right|^2 f_{e0}\left(\frac{\omega}{k}\right) + \frac{2 \pi Z}{k} \left|\frac{\chi_e}{\epsilon}\right|^2 f_{i0}\left(\frac{\omega}{k}\right)
	\end{equation}
	
	where $\vec{k} = \vec{k_s} - \vec{k_i}$ and $\omega = \omega_s - \omega_i$, and $i$ and $s$ denote the incident and scattered electromagnetic radiation waves, corresponding the to probe laser and the Thomson scattered spectrum, respectively.
	
	The total longitudinal dielectric susceptibility is a simple sum of electron and ion components $\epsilon\left(\vec{k}, \omega\right) = 1 + \chi_e\left(\vec{k}, \omega\right) + \chi_i\left(\vec{k}, \omega\right)$. The electron and ion susceptibilities for an unmagnetized, collisionless plasma are described by \cite{sheffield2010plasma}:
	\begin{equation}\label{eq:chiCauchy}
		\chi_{e,i}\left(\vec{k}, \omega\right) = \frac{\omega_{pe,pi}^2}{k^2} \int_{-\infty}^{+\infty} \frac{\vec{k} \cdot \frac{\partial f_{e0, i0}(\vec{v})}{\partial \vec{v}}}{\omega - \vec{k} \cdot \vec{v} - i\gamma} d\vec{v}
	\end{equation}
	where $e$ and $i$ denote electron and ion species respectively, $f$ is the velocity distribution function for the species, and $\omega_p$ is the plasma frequency for the species. This Cauchy integral is taken in the limit as $\gamma \rightarrow 0^{+}$ such that we avoid integrating over the pole by moving into the upper-half of the complex plane, and instead obtain a residue by integrating around the pole. This residue becomes the imaginary component of the susceptibility and it represents the losses due to dissipative processes (Landau damping) in the system.
	
	For a 1D (isotropic) Maxwell-Boltzmann velocity distribution, $f_o \left(\vec{v}\right) = \left(\frac{m}{2 \pi k_B T}\right)^{1/2} \exp \left[- \frac{m v^2}{2 k_B T}\right]$, the dielectric susceptibility reduces to a function which depends directly on the Faddeeva function, $w(x)$:
	
	\begin{equation}\label{eq:chiFaddeeva}
		\chi_{e,i}\left(\vec{k}, \omega\right) = \alpha_{e,i}^2 \left(1 + i \sqrt{\pi} x w(x_{e,i})\right) 
	\end{equation}
	
	The Faddeeva function is numerically implemented in SciPy under the function name \texttt{wofz()}, and, in the future, an implementation of the susceptibility function in Eq \ref{eq:chiFaddeeva} can be found as part of the PlasmaPy project \cite{scipy, plasmapy}. In general, for non-Maxwellian distribution functions the Faddeeva function may not be used and we must numerically evaluate the Cauchy integral in Eq \ref{eq:chiCauchy} instead; non-Maxwellian distributions will not be covered in this article.
	
	The dimensionless phase velocity of the scattering electromagnetic wave is 
	\begin{equation}
		x_{e,i} = \frac{\omega}{\sqrt{2} k v_{The,Thi}}
	\end{equation}
	where $v_{Th}$ is the thermal velocity of the species $v_{The, Thi} = \sqrt{k_B T_{e,i} / m_{e,i}}$.
	
	 The scattering parameter $\alpha$ is 
	\begin{equation}
		\alpha = \frac{\omega_p}{k v_{Th}} = \frac{1}{k \lambda_{De}}
	\end{equation}
	
	where the scattering wavenumber $k$ depends on the scattering angle $\theta$ and the incident and scattered wavenumbers by:
	\begin{equation}
	k = \left(k_s^2 + k_i^2 -2 k_s k_i \cos \theta \right)^{1/2}
	\end{equation}
	
	The geometric meaning of $\theta$ is shown in Figure \ref{fig:gascellschematic}, it is simply the angle between the incident probe beam and the Thomson scattered light measured by the detector. Note that this is the $k$ for a single scattering angle at the probe wavelength. The $k$ which enters into the calculations for $S\left(\vec{k}, \omega\right)$ must be an array of values corresponding to the array of scattered frequencies $\omega_s$ we wish to measure at the detector \cite{sheffield2010plasma}.
	
	The incident and scattered wavenumbers are modulated by the dispersion of the plasma:
	\begin{equation}
		k_i = \frac{\sqrt{\omega_i^2 - \omega_{pe}^2}}{c}
	\end{equation}
	\begin{equation}
		k_s = \frac{\sqrt{\omega_s^2 - \omega_{pe}^2}}{c}
	\end{equation}

	\subsection{Collective and non-collective scattering}
	
	\begin{figure*}
		\centering
		\includegraphics[width=1.0\linewidth]{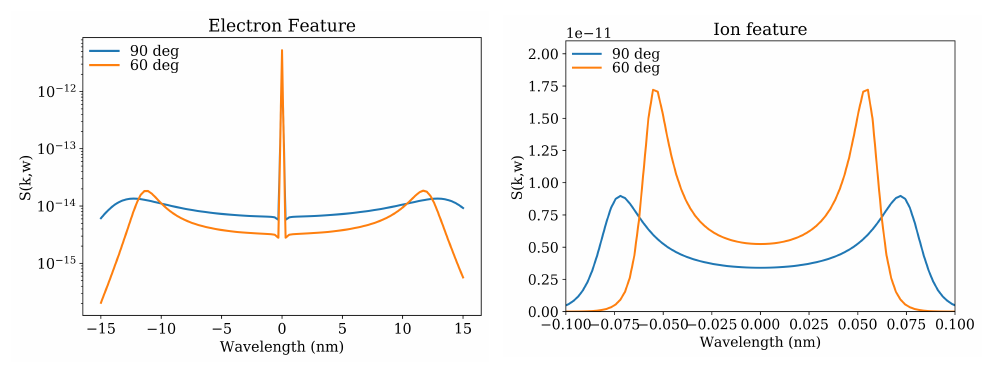}
		\caption{\emph{Left:} Electron feature for Ne plasma with $T_e = T_i = 30 \:\rm eV$, $
			Z = 8$, and $n_e = 1 \times 10^{18} \:\rm cm^{-3}$. The density used here is for illustrative purposes. \emph{Right:} Ion feature for the same conditions. 
			Varying the scattering angle, and therefore $\alpha$ illustrates trends in Thomson scattering spectral features in the collective and non-collective regimes. As $\alpha$ gets smaller and approaches the non-collective scattering regime, the resonances are suppressed and the spectra approach the shape of the distribution function. At $\theta = 90^{\circ}$ we have $\alpha = 1.47$ and $\theta = 60^{\circ}$ corresponds to $\alpha = 2.08$.}
		\label{fig:collectivenoncollectivesubfigures}
	\end{figure*}

	For $\alpha < 1$ we have the non-collective scattering case, where the scale length corresponding to the inverse of the scattering wavenumber is smaller than the screening length of the plasma. This means that scattering will be a superposition of scattering off of each individual particle in the probing volume. The scattered spectrum then effectively reproduces the distribution function of the plasma, because in the limit $\alpha \ll 1$ we have $\chi \rightarrow 0$ and $\epsilon \rightarrow 1$, therefore $S\left(\vec{k}, \omega\right) \rightarrow \frac{2 \pi}{k} f_{e0}\left(\frac{\omega}{k}\right)$. This general trend is demonstrated by simulations in Figure \ref{fig:collectivenoncollectivesubfigures} for a set of illustrative plasma parameters, similar to that of the photoionized Neon experiment, probed by a $2 \omega = 532 \:\rm nm$ laser. Both of these simulations are in the collective regime because the non-collective regime is inaccessible to reasonable scattering angles, under the given plasma and laser conditions. Nonetheless the change in $\alpha$ is significant enough to illustrate the general trend of the EPW resonances smoothing out in favor of the quasi-elastic peak at $\Delta \lambda = 0$, which approaches the shape of the distribution function. 
	
	For $\alpha > 1$ we have the collective scattering case, where the scattering scale length is larger than the screening length of the plasma. In this case the scattering is off of collections of particles, \emph{i.e.}, electron plasma waves, ion-acoustic waves, and other resonances which may be present in the medium. These resonances appear as features within the scattered spectrum at a characteristic upshift/downshift with respect to the probe wavelength. In the left panel of Figure \ref{fig:collectivenoncollectivesubfigures} we see that the electron plasma wave feature is upshifted and downshifted from the central probing frequency (normalized to $\lambda = 0 \:\rm nm$) by $\sim 12-13 \:\rm nm$, depending on scattering angle. The right panel of Figure \ref{fig:collectivenoncollectivesubfigures} zooms in on the pseudo-Rayleigh feature at $0 \:\rm nm$ and is able to resolve this feature as a pair of upshifted/downshifted ion-acoustic wave resonances. The shifts here are $\sim 50-75 \:\rm pm$ away from the central probe wavelength, depending on the scatter angle.
	
	A simple equation for quickly estimating the wavelength separation between the two electron plasma wave features, assuming $\theta = 90^{\circ}$ and $n_e / n_c \leq 0.05$ is given by:
	\begin{equation}
	\frac{\Delta \lambda_{EPW}}{\lambda_i} \approx 2 \left[\frac{n_e}{n_c} + 6 \left(\frac{v_{Th}}{c}\right)^2\right]^{1/2} \left(1 + \frac{3}{2} \frac{n_e}{n_c}\right)
	\end{equation}
	where $n_e$ is the electron density, $n_c = \epsilon_0 m_e \omega_i^2 / q_e^2$ is the critical plasma density of the probe laser, $\Delta \lambda_{EPW}$ is the separation between the electron plasma wave resonances, and $\lambda_i$ is the wavelength of the incident probe laser.
	
	A similar estimate of wavelength separation between the two ion-acoustic wave features is given by:
	\begin{equation}\label{eq:IAWseparation}
	\frac{\Delta \lambda_{IAW}}{\lambda_i} \approx \frac{4}{c} \sin \left(\frac{\theta}{2}\right) \sqrt{\frac{k_B T_e}{m_i} \left(\frac{Z}{1 + k^2 \lambda_{De}^2} + \gamma \frac{T_i}{T_e}\right)}
	\end{equation}
	where $c$ is the speed of light, $m_i$ is the mass of the ion, $Z$ is the average ionization, $T_i$ is the ion temperature and $\gamma$ is the adiabatic index which is typically assumed to be $\gamma = 3$.
	
	By requiring that the second term under the square root in Eq \ref{eq:IAWseparation} be larger than or equal to the first term, we arrive at the criterion for observing the ion-acoustic wave features: $\alpha \gtrsim \left(Z T_e / 3 T_i - 1\right)^{-1/2}$ \cite{ross2010thomson}. 
	Now that we have an initial idea of the expected Thomson scattering features and the constraints for resolving these features, we will examine possible probe laser and spectrometer configurations for implementing optical Thomson scattering on Z.
	
	\section{Diagnostic design}
	
	\subsection{Probe laser}

	In order to cut down on costs for building this new optical Thomson scattering diagnostic on the Z facility, it would be ideal to use equipment which is already available. The typical energy and pulse width ranges for optical Thomson scattering diagnostics applicable to plasmas with $n_e \gtrsim 1 \times 10^{17} \:\rm cm^{-3}$ are $1 - 100 \:\rm J$ and $0.5 - 10 \:\rm ns$ respectively. Fortunately, a suitable laser within this parameter range is already present at the Z facility. The CHACO laser is an Nd:YAG laser, part of the Z-Backlighter Laser Facility, with $16 - 50 \:\rm J$ of energy at a frequency doubled wavelength of $532 \:\rm nm$ over $0.2 - 6 \:\rm ns$ pulse lengths \cite{rambo2016sandia}. In principle the pulse lengths can be extended up to $10 \:\rm ns$, though the authors are not aware of any tests in this operating regime. The CHACO laser also has a special 8-pulse capability with inter-pulse intervals as small as $1 \:\rm ns$, which can be exploited with high speed framing cameras. It should be noted that the maximum energy at the shortest pulse length, that is $0.2 \:\rm ns$, is limited to $16 \:\rm J$. It should also be noted that the CHACO laser is already synchronized to the Z Facility master clock \cite{rambo2016sandia}.
	
	\begin{figure*}
		\centering
		\includegraphics[width=0.90\linewidth]{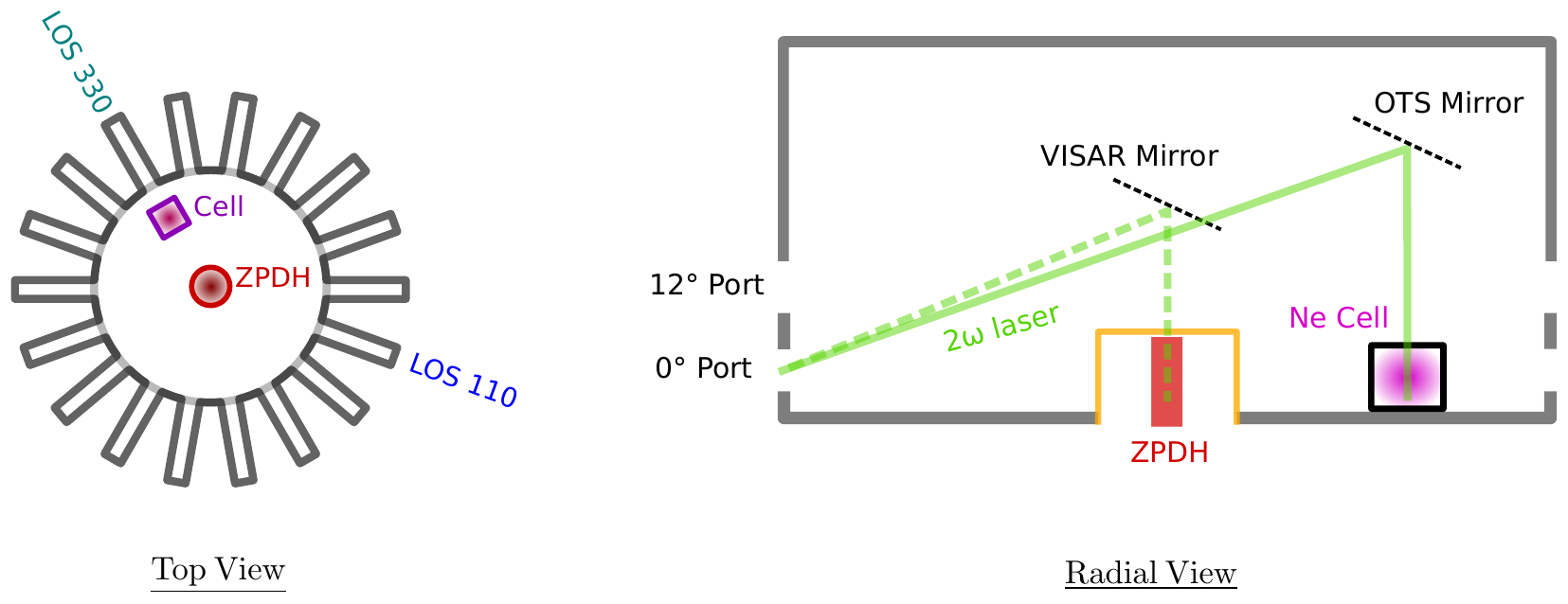}
		\caption{\emph{Left:} Top view of the Z chamber displaying the 18 radial LOS (line of sight) ports. The lines of sight are spaced $20^\circ$ apart and named after their respective angular positions. The TREX diagnostic for the Ne gas cell experiments is positioned to view along LOS 330 (green). The gas cell (purple) sits inside of the chamber at some radial offset from the ZPDH (red). The VISAR beam enters the chamber via LOS 110 (blue). \emph{Right:} Radial view of the Z chamber displaying the $0^\circ$ and $12^\circ$ inclination ports, with respect to the pinch, which exist on each LOS. The VISAR beam (dashed green) will enter via the $0^\circ$ port on LOS 110 and will be reflected vertically into the central pinch region with a disposable mirror. The OTS diagnostic will require a different mirror configuration to reflect the CHACO laser (solid green) into the Ne gas cell (purple). Unlike the VISAR mirror, the OTS mirror does not need to be positioned directly above the pinch, nor does it need to view the pinch, and therefore it can be shielded to reduce costs.}
		\label{fig:lossubfigures}
	\end{figure*}

	The next step is to investigate feasible methods for directing this laser into the Z target chamber and towards the experiment of interest. Co-injection onto an existing beamline would be more cost effective than building a new, dedicated beamline. A velocity interferometer system for any reflector (VISAR) diagnostic is currently being constructed to enter the target chamber through LOS 110 at a $0^{\circ}$ inclination port, as shown in Figure \ref{fig:lossubfigures}. Provided that the optics can support the CHACO laser without being damaged (due to the substantially larger energy requirements for optical Thomson scattering over VISAR), this beamline would be a good candidate for co-injection as the beginning of the beamline sits only a few feet away from the CHACO laser. The VISAR beamline is relay-imaged and under positive air pressure relative to atmosphere. Positive pressure is used to prevent external particulates from entering and contaminating the beamline. Relay imaging requires lenses which bring the VISAR beam to a focus at multiple points along the line. While these two conditions are necessary for VISAR, they pose a problem for co-injection of the CHACO laser for optical Thomson scattering - namely laser-induced breakdown of air. While laser-induced breakdown of air is not an issue for the VISAR laser, as the energy is quite low, it is a concern for the higher energy CHACO laser. There are two obvious ways to mitigate this. The high cost route involves redesigning the beamline to be placed under vacuum, such that there is insufficient air to absorb laser energy and cause breakdown. The low cost route involves removing the relay lenses while propagating the CHACO laser along the beamline, such that it never comes to a focus.
	
	Note that a significant cost to fielding the VISAR diagnostic is the use of in chamber disposable optics. This is because the VISAR line is used to observe back reflected laser light from the central pinch, as shown in Figure \ref{fig:lossubfigures}. Therefore the axial radiation and plasma from the pinch will destroy those optics. We have no such requirements for optical Thomson scattering, both because we are not probing the pinch and because we are measuring scattering instead of back reflection. Therefore we can position and shield our optics to mitigate damage caused by the pinch, and keep costs low.
	
	An alternative to co-injection into the VISAR beamline would be to co-inject into the Z Beamlet (ZBL) beamline. This beamline also supports $2 \omega$ light, but at much higher energies, and enters the target chamber axially. Unfortunately, this would restrict the axial line of sight from being used for the Fe opacity experiments which are fielded as part of ZAPP. Axially fielding the CHACO laser may also generate higher costs if disposable optics need to be fielded directly above the pinch to redirect the laser to the experiment under test.
	
	The primary form of laser induced damage, for which we need to estimate tolerable energy density thresholds of the VISAR optics, is expected to be dielectric breakdown. This is based on the relatively short pulse length of the CHACO laser \cite{wood2003laser}. As a follow-up to this study, detailed calculations on laser induced damage thresholds will have to be conducted to verify that the CHACO laser can be safely propagated along the VISAR beamline.
	
	\subsection{Collection optics}

	The solid angle of the collection optic relative to the plasma is taken to be $\Omega = 6 \times 10^{-5}\:\rm sr$. This is the measured solid angle for a separate gas cell system which is studying conditions related to white dwarf photospheres and is fielded as part of the ZAPP collaboration \cite{falcon2013experimental,falcon2015laboratory}. The gas cell size is comparable to the Ne gas cell and uses the SVS systems to measure optical emission spectra, therefore it is reasonable as an initial estimate for the types of collection optics which could be implemented to measure Thomson scattered spectra. The one caveat here is that the white dwarf gas cell uses $3 \:\rm mm$ apertures to restrict radiation from a gold back wall within the cell from being collected, and this ultimately limits the solid angle subtended by the collection optics. Although this gold back wall is intrinsic to the design of the white dwarf gas cell experiments, it is not found in the Neon gas cell experiments and therefore the aforementioned solid angle should be taken as a conservative estimate for what is possible.
	
	The collected light must subsequently be coupled to a fiber optic line to be transported to the SVS spectrometer systems, approximately $\sim 70 \:\rm m$ away. Both the coupling of the collection optic with the fiber optic line, and length of the line itself cause losses in signal. The collection optic for the white dwarf experiment is coupled to a single optical fiber with a $100 \:\rm \mu m$ core diameter. It is assumed that a similar collection system would be implemented for measuring OTS signals on the Neon gas cell experiment. A conservative estimate of typical losses due to coupling, limiting apertures, windows, solid angles and the length of optical fiber yields a transmission of $\sim 10 \: \%$; this is based on calibration measurements from the white dwarf photosphere experiments. The losses along the length of the optical fiber itself yield a transmission of $\sim 50 \%$, Therefore substantial signal losses can be mitigated by using a simpler collection setup than that of the white dwarf photosphere experiments as well as by using a spectrometer closer to the target chamber.

	\subsection{Spectrometer response}

	The streaked visible spectroscopy (SVS) system consists of 5 spectrometers, labeled SVS1 through SVS5. These spectrometers have a wide range of gratings available, from $150 \:\rm \ell/mm$ to $2400 \:\rm \ell/mm$ which are suitable for resolving both the electron and ion features. The fastest streak is $100 \:\rm ns$ for the CCD based spectrometers and $25 \:\rm ns$ for the film based spectrometers. The latter streak length may be suitable for time-resolved measurement of optical Thomson scattering spectra using the CHACO laser with a longer pulse width as the probe. This kind of temporal resolution would allow for tracking of the evolution of plasma parameters before, during, and after the $\sim 3 \:\rm ns$ of peak x-ray flux (and photoionization) characteristic of the ZPDH \cite{rochau2014zapp}. The time-resolved plasma parameters from a streaked OTS diagnostic may then also be compared with the time-resolved x-ray spectroscopic measurements provided by the TREX diagnostic.
	
	It is important to know the detection threshold and signal to noise ratio inherent to the SVS systems (due to effects such as dark current) for given input powers, so that we may check whether the radiation power of the Thomson scattered signal incident on the spectrometer is measurable.
	
	SVS5, a CCD-based spectrometer was tested with a $600 \:\rm \ell/mm$ grating. This grating is what would be used to resolve the electron plasma wave feature, which will have the lowest signal level compared to other features in the Thomson scattering spectrum. The spectrometer was used to record a laser for a range of powers (measured using a calibrated power meter) and with a wavelength of $544 \:\rm nm$ due to the lack of a stable, low power $532 \:\rm nm$ laser. It should be noted that the response of the spectrometer is quite flat over this wavelength range, so there should not be any significant differences in the detector response between $544 \:\rm nm$ and $532 \:\rm nm$. The dispersed laser was streaked over $100 \:\rm ns$ and recorded onto the CCD.
	
	The recorded laser spectra were then processed by fitting a Gaussian to the laser profile. This Gaussian is primarily due to the instrument function of the spectrometer and could be convolved with the electron feature simulations presented in subsequent sections.
	
\begin{figure*}
		\centering
		\includegraphics[width=0.9\linewidth]{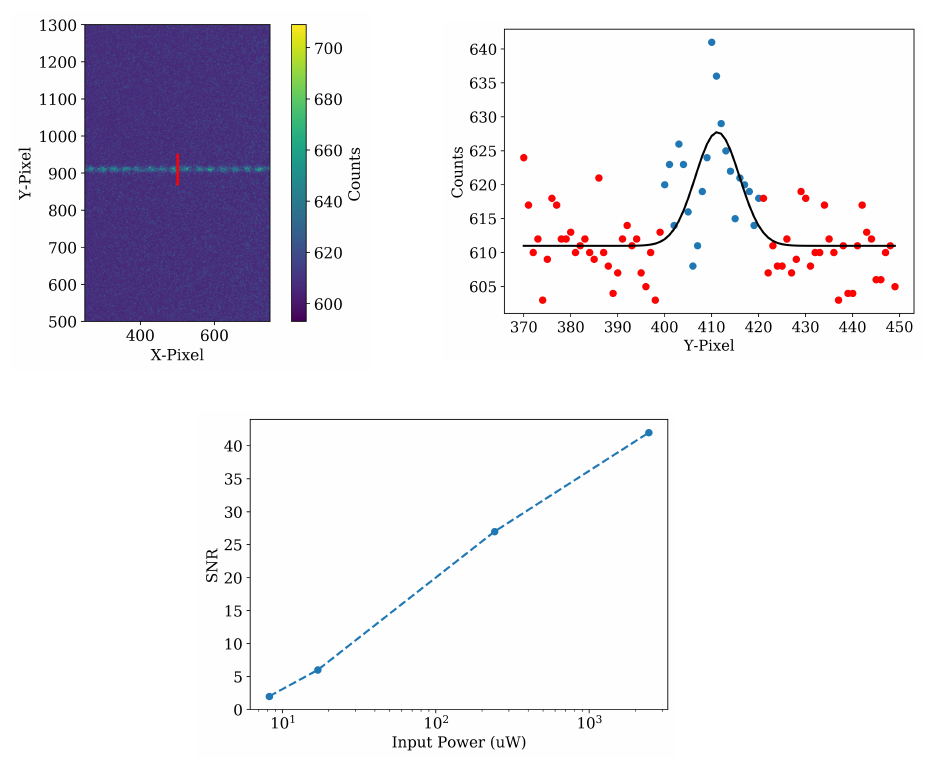}
		\caption{\emph{Top Left:} Streaked image of laser at $544 \:\rm nm$ recorded on SVS5. The horizontal axis is time resolution and the vertical axis is spectral resolution. A red line, 1 px thick in the temporal direction represents the lineout region. Note that the laser appears as a sequence of spots due to modulations in the laser power. The lineout is taken in a region between the brightest and darkest spots such that it is representative of the average laser power. \emph{Top Right:} An example of lineout data (points) with Gaussian fit (black) used to obtain signal to noise ratio for SVS5. Red points represent noise, which is taken to be data outside of $5 \sigma$, while blue points are signal. For this particular fit we have $\rm SNR \sim 2$. \emph{Bottom:} The signal to noise response of SVS5 taken across multiple laser input powers.}
		\label{fig:svssnrresponse}
	\end{figure*}

	We calculate the signal to noise as the ratio between the mean of the signal amplitude (the fitted Gaussian within $\pm 5\sigma$ along the spectral axis) and the standard deviation of the noise amplitude (the measured counts outside the $\pm 5\sigma$ of the fitted Gaussian.)
	\begin{equation}
		{\rm SNR} = \frac{\mu_{\rm signal}}{\sigma_{\rm noise}}
	\end{equation}
	The processing steps for these SVS calibration measurements and the resultant SNR of SVS5 for various input powers is summarized in Figure \ref{fig:svssnrresponse}. We see that even for the relatively low input power of $10 \:\rm \mu W$, the detector's signal to noise is $\rm SNR \sim 2$ when integrating across a single pixel in the temporal axis. Of course, the signal to noise can be further increased by integrating across additional pixels in the temporal axis, thereby sacrificing temporal resolution.
	
	\section{Simulations}
	
	In this section we will run through a number of simulations which set constraints on our diagnostic requirements. The noise produced by background Bremsstrahlung will determine the lower limit on the probe laser power and whether we need to use a polarized laser and a polarization filter on the spectrometer to boost the signal to noise ratio of the Thomson scattered spectra. Radiation hydrodynamic simulations will check for temperature perturbations due to heating by the probe laser, and will set the upper limit on the probe laser power. The optical Thomson scattering diagnostic must then be designed within the lower and upper bounds of these laser power constraints. Furthermore, detailed simulations of the Thomson scattered spectra while varying temperature and density independently will set constraints on the required spectrometer resolution (instrument function) of the laser-spectrometer system.
	
	\subsection{Bremsstrahlung radiation power}
	
	The primary sources of background noise which may interfere with the measurement of Thomson scattering spectra, are due to plasma emission, whether as line emission or Bremsstrahlung. We therefore estimate the radiation power in Bremsstrahlung emission from photoionized Ne since this may contribute substantially to the background noise against which we are measuring the Thomson scattered signal.
	
	The Bremsstrahlung power in Watts is given by \cite{sheffield2010plasma}:
	\begin{equation}
	\begin{split}
	P_B d \Omega d \lambda_s = 2.09 \times 10^{-36} g Z^2 \left(\frac{n_e n_i}{\lambda_s^2 T_e^{1/2}}\right) \\
	\times \exp\left[-\left(\frac{1.24 \times 10^{-4}}{\lambda_s T_e}\right)\right] V_p \frac{d \Omega}{4 \pi} d \lambda_s
	\end{split}
	\end{equation}
	where $P_B$ is the Bremsstrahlung power, $d\Omega$ is the differential solid angle, $\lambda_s$ is the wavelength of the Bremsstrahlung radiation in units of $\rm cm$, $Z$ is the average ionization of the plasma, $g$ is the Gaunt factor, $n_e$ and $n_i$ are the electron and ion densities in units of $\rm cm^{-3}$, $V_p$ is the volume over which we are collecting Bremsstrahlung radiation in units of $\rm cm^3$, and $T_e$ is the electron temperature in units of $\rm eV$.
	
	Assuming the range of angles over which Bremsstrahlung is measured is relatively small, and discretizing the scattered wavelength $\lambda_s$ we simplify the Bremsstrahlung power to
	
	\begin{equation}\label{eq:bremsstrahlung}
	\begin{split}
		P_B = 2.09 \times 10^{-36} g Z^2 \left(\frac{n_e n_i}{\lambda_s^2 T_e^{1/2}}\right) \\
		\times \exp\left[-\left(\frac{1.24 \times 10^{-4}}{\lambda_s T_e}\right)\right] V_p \frac{\Omega}{4 \pi} \lambda_s
	\end{split}
	\end{equation}
	
	We simulate the Bremsstrahlung radiation over just the wavelength range of interest to Thomson scattering. Here we have assumed a Gaunt factor of order unity $g \sim 1$. As mentioned previously, we assume a solid angle of $\Omega = 6 \times 10^{-5} \:\rm sr$ for the collection optic, based on the white dwarf photosphere experiments. We take the volume over which we are collecting Bremsstrahlung emission as $V_p = A l_p$, where $A$ is the spot size of the probe laser and $l_p$ is the length of the probed volume, which is constrained by the collection optic. The spot area is given by $A = \pi r_p^2$ where $r_p$ is the radius of the laser spot. We will assume a spot radius of $r_p = 500 \:\rm \mu m$ and probe length of $l_p = 5 \:\rm mm$, which is about a fifth of the longest dimension of the gas cell, providing reasonable spatial resolution, while maximizing the collected Thomson scattering signal. The expected Bremsstrahlung radiation from our scattering volume is relatively flat in the spectral region of interest for Thomson scattering with a $2 \omega$ probe and is about $\sim 17 \:\rm mW$ of radiated power for $n_e = 4 \times 10^{18} \:\rm cm^{-3}$ and $\sim 68 \:\rm mW$ of radiated power for $n_e = 8 \times 10^{18} \:\rm cm^{-3}$, at the Thomson scattering collection optic.
	
	As the Bremsstrahlung is randomly polarized, whereas Thomson scattering may be polarized given a polarized probe beam, the signal to noise ratio between Thomson scattered radiation and Bremsstrahlung radiation may be increased by introducing a polarizer on the collection line \cite{sheffield2010plasma}.
	
	\subsection{Thomson scattered radiation power}
	
	The next step is to estimate the expected power in Thomson scattered signal at our collection optic, so that we may compare it to the background noise as well as to the sensitivity of our spectrometer.
	The optical Thomson scattering power is given by:
	\begin{equation}\label{eq:otsPower}
	\begin{split}
	P \left(\vec{R}, \omega_s\right) d\Omega d\omega_s = \frac{P_i r_0^2}{A 2 \pi} d\Omega d\omega_s \\
	\left|\hat{s} \times \left(\hat{s} \times \hat{E}_{i0}\right)\right|^2 N S\left(\vec{k}, \omega\right)  
	\end{split}
	\end{equation}
	where $P_i$ is the power of the laser probe, $r_0 = e^2 / 4 \pi \epsilon_0 m_e c^2 \approx 2.818 \times 10^{-15} \:\rm m$ is the classical electron radius, $A$ is the probe spot size, $N = n_e V_p = n_e A l_p$ is the total number of scatterers in the cylindrical probe volume.
	
	For a polarized probe \cite{sheffield2010plasma}
	\begin{equation}
	\left|\hat{s} \times \left(\hat{s} \times \hat{E}_{i0}\right)\right|^2 = 1 - \sin^2 \theta \cdot \cos^2 \phi_0
	\end{equation}
	and for an unpolarized probe \cite{sheffield2010plasma}
	\begin{equation}
	\left|\hat{s} \times \left(\hat{s} \times \hat{E}_{i0}\right)\right|^2 = 1 - \frac{1}{2} \sin^2 \theta
	\end{equation}
	
\begin{figure}
	\centering
	\includegraphics[width=1.0\linewidth]{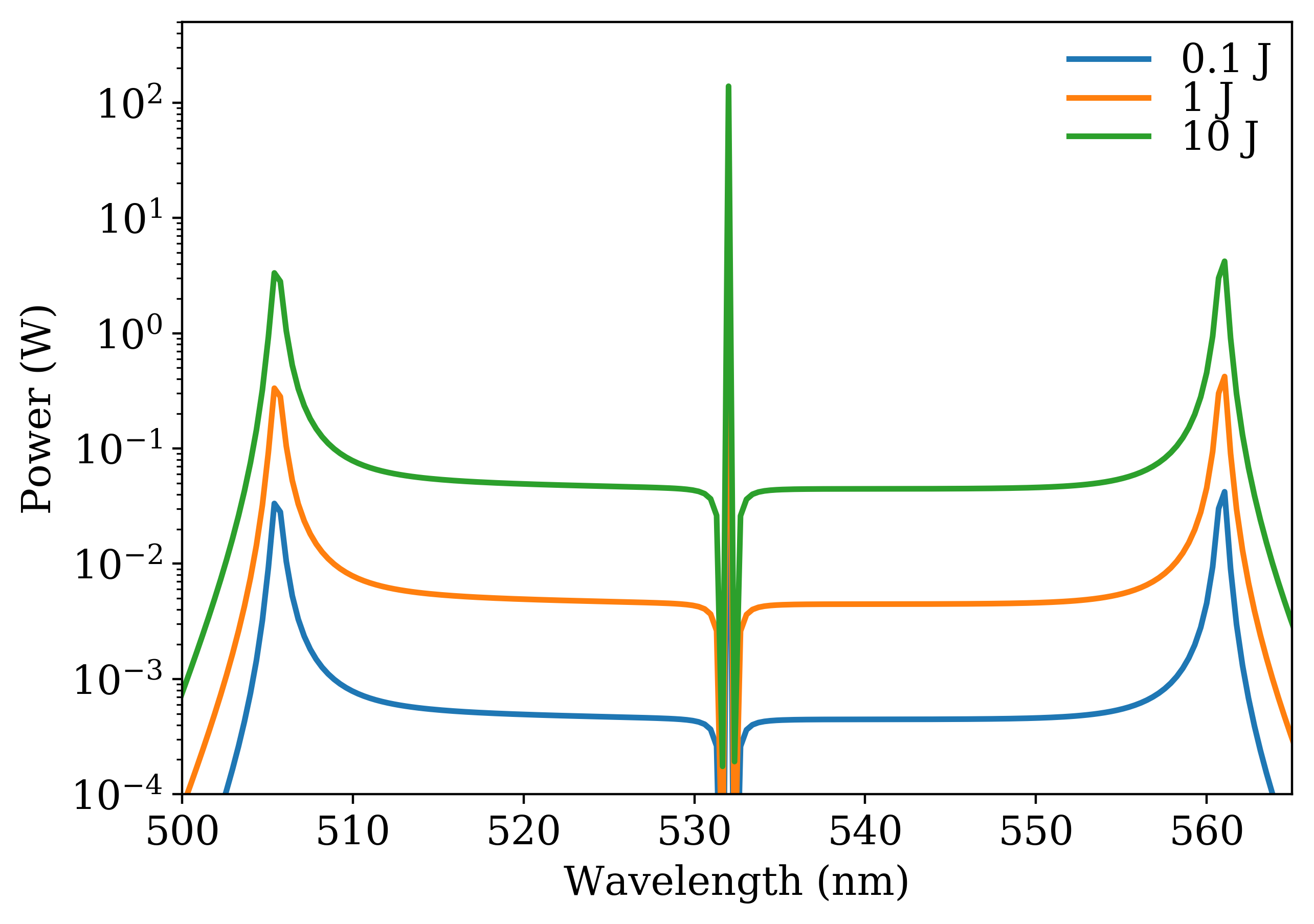}
	\caption{Electron plasma wave and pseudo-Rayleigh scattering features for different laser energies and pulse widths. The power is that incident on the collection optic. The plasma conditions for these simulations were: $n_e = 8 \times 10^{18} \:\rm cm^{-3}$, $T_e = T_i = 30 \:\rm eV$, $Z = 8$. The probe conditions were: $\theta = 120^{\circ}$, $\lambda_i = 532 \:\rm nm$, $\Delta t = 1 \:\rm ns$, $r_p = 500 \:\rm \mu m$, $l_p = 5 \:\rm mm$, $\Omega= 6 \times 10^{-5} \:\rm sr$.}
	\label{fig:otspowerplot}
	\end{figure}

	It is evident from Eq. \ref{eq:otsPower} that the amount of Thomson scattering radiation available for measurement is highly dependent on the probe power and the number of scatterers available in the probe volume. It is important to verify that the amount of Thomson scattered signal for a given configuration is measurable by the spectrometer. We simulate the radiated Thomson scattering power for a range of laser energies: $0.1 - 10 \:\rm J$ with fixed pulse width $\Delta t = 1 \:\rm ns$. We use the following parameters (consistent with the Bremsstrahlung calculation): a spot radius of $r_p = 500 \:\rm \mu m$, probe length of $l_p = 5 \:\rm mm$, collection solid angle of $\Omega = 6 \times 10^{-5} \:\rm sr$, probe wavelength $\lambda = 532 \:\rm nm$, scatter angle $\theta = 120^{\circ}$ for a plasma at $n_e = 8 \times 10^{18} \:\rm cm^{-3}$, $T_e = T_i = 30 \:\rm eV$, $Z = 8$. The results of these simulations are summarized in Figure \ref{fig:otspowerplot}. We see that the power at the collection optic is $\sim 30 \:\rm mW$ for the lowest power signal, that is the electron plasma wave feature, using an unpolarized probe laser. Using the estimated losses along the optical fiber, due to transporting the scattered radiation from the collection optics in the target chamber to the spectrometer, we have $\sim 3 \:\rm mW$ of power due to Thomson scattered radiation incident at the SVS spectrometer. This is well above the detection threshold, as seen in Figure \ref{fig:svssnrresponse} and a good signal-to-noise ratio is expected.
	
	Furthermore, we must compare Thomson scattered power to background noise from the plasma, \emph{i.e.}, Bremsstrahlung. We expect $30 \:\rm mW$ due to Thomson scattered radiation for the lowest laser energy we simulated, which is about half of $\sim 68 \:\rm mW$ of radiation power due to Bremsstrahlung at the collection optic. Therefore, signal to noise levels will be a significant challenge when using $0.1 \:\rm J$ of laser energy and a polarized laser with polarizing filter on the spectrometer will have to be used to boost the signal to noise at this energy level. For the $1 \:\rm J$ laser case it is clear that we have $\sim 200 \:\rm mW$, which is well above the Bremsstrahlung noise level.
	
	We neglect the background radiation from the ZPDH incident on the gas cell, as the power from the pinch near $532 \:\rm nm$ is expected to be quite small. This is because the pinch temperature peaks at $\sim 300 \:\rm eV$, resulting in a peak in Planckian emission spectra in the ultraviolet wavelength range.
	
	Finally, we must check for possible emission lines near $532 \:\rm nm$ which may interfere and blend with the Thomson scattering spectra. A simple lookup of the NIST tables for spectral lines emitted by Ne shows that only Ne I and Ne II have sufficiently strong oscillator strengths in the $532 \pm 30 \:\rm nm$ range, but at $30 \:\rm eV$ temperatures we do not expect to have significant amounts of these neutral and low ionization states present within the gas cell \cite{kramida2018nist}. A detailed accounting of other materials which may cause line emission along the line of sight will have to be conducted in the future.
	
	\subsection{Electron feature}
	
	\begin{figure*}
		\centering
		\includegraphics[width=1.0\linewidth]{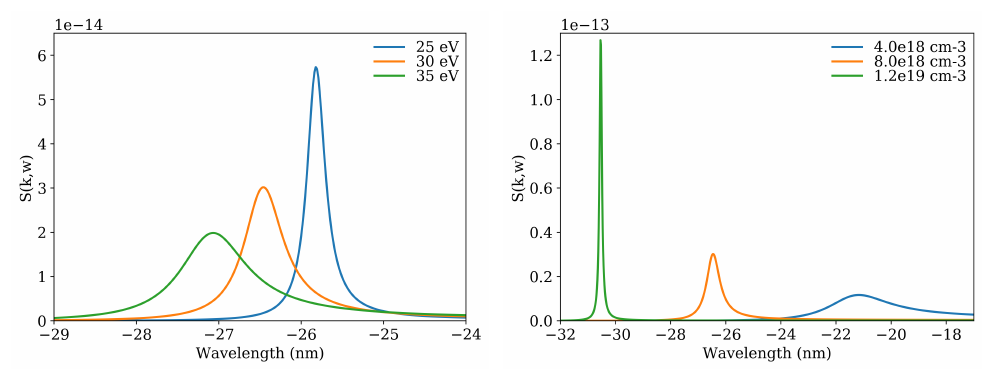}
		\caption{\emph{Left:} Sensitivity of downshifted EPW feature to variations in electron temperature while keeping all other plasma parameters (including ion temperature) fixed. \emph{Right:} Sensitivity of downshifted EPW feature to electron density while scaling ion density by average ionization.}
		\label{fig:epwsensitivitysubfigures}
	\end{figure*}
	
	We vary the plasma electron temperature and density while keeping the scattering geometry fixed to test the sensitivity of the Thomson scattering diagnostic, as well as the necessary resolution for our spectrometers. All simulations were run for $n_e = 8.0 \pm 4.0 \times 10^{18} \:\rm cm^{-3}$, $n_i = n_e / Z$, $Z = 8$, $T_e = 30 \pm 5 \:\rm eV$, $T_i = 30 \:\rm eV$, $\theta = 120^{\circ}$, and $\lambda_i = 532 \:\rm nm$ . The results of these simulations, focusing on the downshifted EPW feature, are displayed in Figure \ref{fig:epwsensitivitysubfigures}. We see that the shift is $\Delta \lambda_{\rm EPW} \approx 26 \:\rm nm$, which sets a limit on the instrument function to prevent the ion and electron features from overlapping so that we may resolve the positions of the resonances. The peak widths are $\Delta \lambda \approx 1 \:\rm nm$ which further constrains the requisite resolution of the spectrometer in the case where we want to resolve the shape of the EPW feature. We want information on both the position and shape of the resonances to better constrain inferred plasma parameters, and therefore we must use the stricter resolution constraint.
	
	\subsection{Ion feature}
	
	\begin{figure*}
		\centering
		\includegraphics[width=1.0\linewidth]{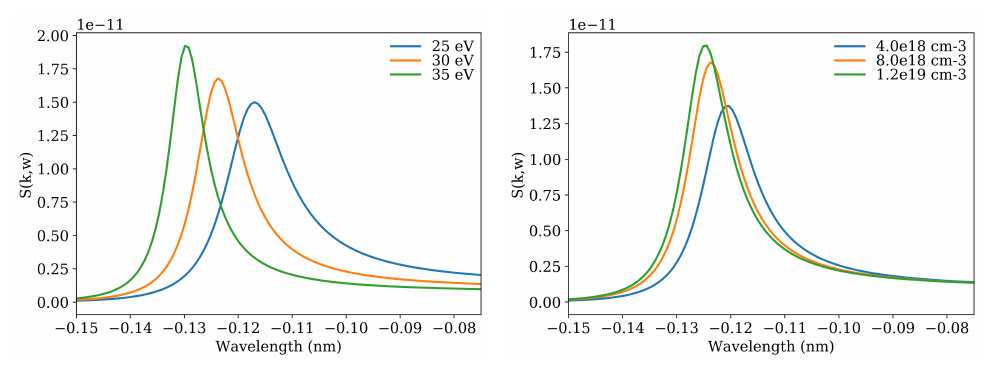}
		\caption{\emph{Left:} Sensitivity of downshifted IAW feature to variations in electron temperature while keeping all other plasma parameters (including ion temperature) fixed. \emph{Right:} Sensitivity of downshifted IAW feature to electron density while scaling ion density by average ionization.}
		\label{fig:iawsensitivitysubfigures}
	\end{figure*}
	
	Using the same plasma and probe parameters as for the EPW feature, we test the sensitivity of the IAW feature to temperature and density changes. The results of these simulations, focusing on the downshifted IAW feature, are displayed in Figure \ref{fig:iawsensitivitysubfigures}.  We see that the shift is $\Delta \lambda_{\rm IAW} \approx 120 \:\rm pm$, which sets the limit on the instrument function and resolution to prevent the upshifted and downshifted IAW features from blending together.

	\subsection{Probe laser perturbation}
	One significant source of perturbation is due to individual particles in the plasma being accelerated by the probe laser's electric field. The criterion for this form of perturbation tests the velocity of an electron in the laser E-field compared to its thermal velocity. The velocity due to this E-field should be insignificant compared to the thermal velocity, such that the distribution of velocities is undisturbed.
	To assess this effect, we use the following inequality \cite{sheffield2010plasma}
	\begin{equation}\label{eq:perturb}
		\frac{P_i}{A} \ll \frac{c \epsilon_0 m_e^2 v_{Th}^2}{2 q_e^2}\omega_i^2
	\end{equation}
	where we assume that the laser is significantly perturbing the plasma if the left hand side of Eq \ref{eq:perturb} is $\geq 1 \: \%$ of the right hand side.
	
	For our $30 \:\rm eV$ temperature Ne plasma, probed by a $532 \:\rm nm$ laser, the right hand side is $\sim 5.7 \times 10^{18} \:\rm W/m^2$. Using the highest power configuration for CHACO, with a laser energy of $16 \:\rm J$ over $200 \:\rm ps$ and a spot radius of $500 \:\rm \mu m$ the left hand side is $\sim 1.0 \times 10^{17} \:\rm W/m^2$, which is $\sim 1.75 \: \%$ of the right hand side of Eq \ref{eq:perturb}. For a laser energy of $10 \:\rm J$ over $1 \:\rm ns$ and a spot radius of $500 \:\rm \mu m$ the left hand side is $\sim 1.0 \times 10^{16} \:\rm W/m^2$, which is $\sim 0.18 \: \%$ of the right hand side of Eq \ref{eq:perturb}. Therefore, the probe laser should be operated away from the high power regime and towards this intermediate power regime to avoid perturbation of the particle velocities. 
	
	\begin{figure}
		\centering
		\includegraphics[width=1.0\linewidth]{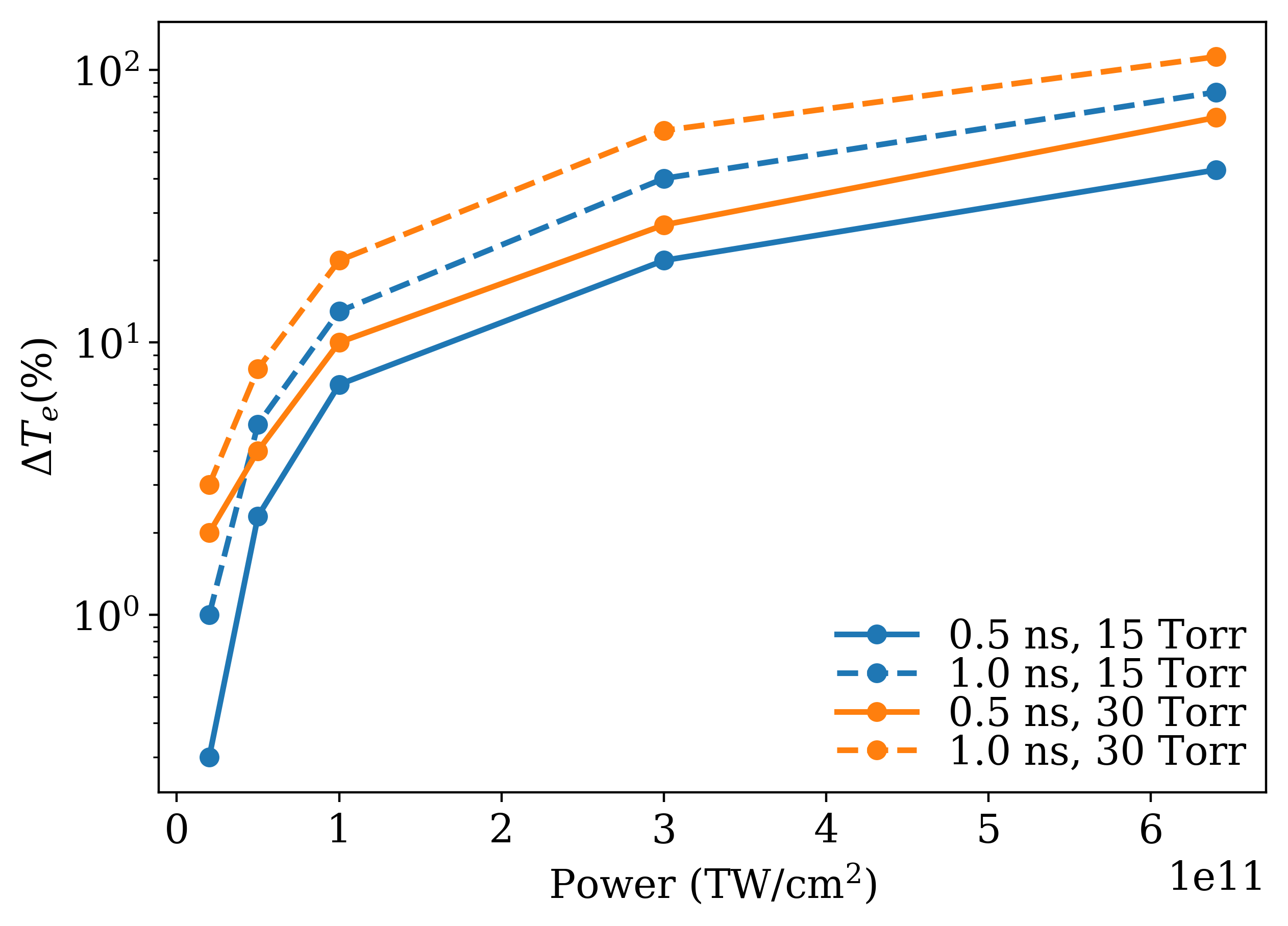}
		\caption{Summary of HELIOS radiation-hydrodynamic simulation results showing expected temperature perturbation due to the Thomson scattering probe beam for a range of incident powers. The $P = 15 \:\rm Torr$ and $P = 30 \:\rm Torr$ cases are displayed in blue and orange respectively. The percentage change in electron temperature is displayed at $0.5 \:\rm ns$ (solid line) and $1.0 \:\rm ns$ (dashed line) after the start of the probe laser.}
		\label{fig:heliostemperaturepercent}
	\end{figure}

	Another significant form of perturbation is heating of the plasma due to the energy deposited by the probe laser. We checked this point by running radiation-hydrodynamic simulations of the plasma under a range of incident probe laser parameters with HELIOS \cite{macfarlane2006helios}. The simulations were done for a Neon slab plasma initialized at $T_e = 30 \:\rm eV$, $n_a = 5 \times 10^{17} \:\rm cm^{-3}$ and $n_a = 1 \times 10^{18} \:\rm cm^{-3}$, and with length of $10 \:\rm mm$. The physics model included the possibility of different electron and ion temperatures, photon energy dependent radiation transport, and LTE as well as NLTE inline atomic physics. The laser wavelength was $532 \:\rm nm$, \emph{i.e.}, green light, it had a $1 \:\rm ns$-squared duration pulse shape, and it was absorbed in the plasma via inverse Bressmtrahlung. This pulse length would provide a reasonable snapshot of the photoionized plasma condtions via Thomson scattering, \emph{i.e.}, electron temperature. 
	
	The goal of these simulations was to estimate the increase in electron temperature due to laser energy deposition. To this end, calculations were performed for both atom number densities, which correspond to gas cell experiments performed with filling pressures $P=15 \:\rm Torr$ and $P=30 \:\rm Torr$ and laser intensities in the range from {$0.01 \:\rm TW/cm^2$} through {$0.6 \:\rm TW/cm^2$}. The results from these simulations are summarized in Figure \ref{fig:heliostemperaturepercent}, and they indicate that the probe laser intensity should be less than {$0.1 \:\rm TW/cm^2$} for the electron temperature increase to be of order {$10\%$}. Therefore, for a spot radius of $500 \:\rm \mu m$ and pulse width of $1 \:\rm ns$ the laser energy should not exceed $\sim 0.8 \:\rm J$. Since the minimum laser energy, as constrained by noise due to Bremsstrahlung, is $\sim 0.1 \:\rm J$ over $1 \:\rm ns$, we can feasibly implement an optical Thomson scattering measurement for the photoionzed Neon gas cell experiments.

	\section{Summary}
	In this initial study we have demonstrated that it is indeed feasible to use optical Thomson scattering as an electron temperature diagnostic for photoionized Neon experiments on Z. Based on our simulations, it is possible to tune the CHACO laser down to the appropriate energy level for simultaneously mitigating heating by the probe and maximizing signal, such that it can be measured by the SVS spectrometer systems.
	
	It should be noted that our estimates throughout this feasibility study have been generally conservative. We set the condition that the OTS signal should be larger than the noise due to Bremsstrahlung, yet this condition need not be so strict. The spectral signal due to Bremsstrahlung follows a curve which is distinct from the OTS signal, and it can often be fitted and removed in post-processing because the OTS and Bremsstrahlung spectra are additive. We were also conservative about signal collection and transmission - primarily relying on the white dwarf gas cell setup. The white dwarf gas cell uses apertures to restrict radiation from a gold back wall within the cell from being collected, and these apertures ultimately restrict the solid angle for signal collection. The Neon gas cell has no such backlighter and therefore there is more room to expand the solid angle subtended by the collection optics, if needed. Furthermore, the SVS spectrometers are located $\approx 70 \:\rm m$ away from the pinch, causing significant losses along the transport fiber optics. These losses can be readily reduced by using a time-gated or spatially resolved spectrometer located in the LOS hutches, just a few meters away from the pinch. Finally, all of our estimates were using the EPW feature as the signal because it is the lowest intensity feature, yet even in the absence of resolving the EPW feature the IAW feature can be used to obtain some information about the plasma. 
	
	On the other hand, there were a few points in our analysis which were optimistic. For one, we estimated the Bremsstrahlung as primarily coming from the Thomson scattering volume, even though plasma just outside of this volume would only be slightly out of focus and could contribute to the overall noise. In addition, no instrument function was convolved with the Thomson scattered spectra, which would have the effect of broadening and lowering the peaks to some extent. Removing the Bremsstrahlung feature in post-processing, and increasing the contrast between OTS and Bremsstrahlung signals by using a polarizer should be sufficient to compensate for the places where our approximations overestimate the signal and underestimate the noise.
	
	Of course, a new optical Thomson scattering diagnostic on Z would by no means be restricted to the study of photoionized Neon plasmas, but could find a wide range of applicability. The photoionized Silicon experiments in the ZAPP collaboartion, for example, have a similar range of plasma densities, and therefore may be suitable for probing by Thomson scattering. It should be noted that application of Thomson scattering for probing plasmas near the central pinch would require additional protection of the optics lines to prevent blanking \cite{swadling2017initial}.
	
	\section{Future work}
	Prior to the implementation of this diagnostic on the Neon gas cell experiments, a number of analyses will have to be conducted on top of the foundations we have laid out in this paper. First, a suitable material for the gas cell window must be selected to allow propagation of the optical Thomson scattering laser. The ablation of this window under simultaneous heating from the Z-pinch X-rays and probe laser must be simulated and used to assess the attenuation of the probe laser through the window plasma. Second, a detailed laser induced damage threshold analysis will have to be conducted for the CHACO laser propagating along the VISAR beamline. Third, a list of materials, and corresponding line emission, present along the Thomson scattering spectrometer line of sight must be compiled to ensure no background signals will interfere with the diagnostic. Last, similar feasibility studies will have to be conducted for a range of plasmas under investigation on Z to demonstrate a wide range of applicability and motivate funding for this diagnostic. In particular, Thomson scattering may find applications in other ZAPP collaboration experiments, such as  supplementing spectroscopic measurements on the photoionized Silicon experiments, or on magnetized liner inertial fusion (MagLIF) experiments as a magnetic field probe.

	\begin{acknowledgments}
		Thanks to David Bliss for assistance in obtaining spectrometer calibrations and for discussions on laser setup. A special thanks to Rachel Hopkins and Ted Lane for assistance in proof reading and informative discussions.
		Thanks to Daniel Mayes for providing expected plasma parameters for the Ne gas cell experiment. Thanks to Michael Jones for providing specifications for the new line VISAR system. Thanks to Marc Schaeuble for providing collection optic information from the white dwarf photosphere experiments.
		
		This work was performed as part of Sandia's Z Fundamental Science Program, and the Z Astrophysical Plasma Properties collaboration at Sandia.
		Financial support from DOE-SC-FES grant DE-SC0012515 is gratefully acknowledged.
	\end{acknowledgments}
	
	\bibliography{OTSonZPaper}

\begin{thebibliography}{19}%
\makeatletter
\providecommand \@ifxundefined [1]{%
 \@ifx{#1\undefined}
}%
\providecommand \@ifnum [1]{%
 \ifnum #1\expandafter \@firstoftwo
 \else \expandafter \@secondoftwo
 \fi
}%
\providecommand \@ifx [1]{%
 \ifx #1\expandafter \@firstoftwo
 \else \expandafter \@secondoftwo
 \fi
}%
\providecommand \natexlab [1]{#1}%
\providecommand \enquote  [1]{``#1''}%
\providecommand \bibnamefont  [1]{#1}%
\providecommand \bibfnamefont [1]{#1}%
\providecommand \citenamefont [1]{#1}%
\providecommand \href@noop [0]{\@secondoftwo}%
\providecommand \href [0]{\begingroup \@sanitize@url \@href}%
\providecommand \@href[1]{\@@startlink{#1}\@@href}%
\providecommand \@@href[1]{\endgroup#1\@@endlink}%
\providecommand \@sanitize@url [0]{\catcode `\\12\catcode `\$12\catcode
  `\&12\catcode `\#12\catcode `\^12\catcode `\_12\catcode `\%12\relax}%
\providecommand \@@startlink[1]{}%
\providecommand \@@endlink[0]{}%
\providecommand \url  [0]{\begingroup\@sanitize@url \@url }%
\providecommand \@url [1]{\endgroup\@href {#1}{\urlprefix }}%
\providecommand \urlprefix  [0]{URL }%
\providecommand \Eprint [0]{\href }%
\providecommand \doibase [0]{http://dx.doi.org/}%
\providecommand \selectlanguage [0]{\@gobble}%
\providecommand \bibinfo  [0]{\@secondoftwo}%
\providecommand \bibfield  [0]{\@secondoftwo}%
\providecommand \translation [1]{[#1]}%
\providecommand \BibitemOpen [0]{}%
\providecommand \bibitemStop [0]{}%
\providecommand \bibitemNoStop [0]{.\EOS\space}%
\providecommand \EOS [0]{\spacefactor3000\relax}%
\providecommand \BibitemShut  [1]{\csname bibitem#1\endcsname}%
\let\auto@bib@innerbib\@empty
\bibitem [{\citenamefont {Harvey-Thompson}\ \emph {et~al.}(2012)\citenamefont
  {Harvey-Thompson}, \citenamefont {Lebedev}, \citenamefont {Patankar},
  \citenamefont {Bland}, \citenamefont {Burdiak}, \citenamefont {Chittenden},
  \citenamefont {Colaitis}, \citenamefont {De~Grouchy}, \citenamefont {Doyle},
  \citenamefont {Hall} \emph {et~al.}}]{harvey2012optical}%
  \BibitemOpen
  \bibfield  {author} {\bibinfo {author} {\bibfnamefont {A.}~\bibnamefont
  {Harvey-Thompson}}, \bibinfo {author} {\bibfnamefont {S.}~\bibnamefont
  {Lebedev}}, \bibinfo {author} {\bibfnamefont {S.}~\bibnamefont {Patankar}},
  \bibinfo {author} {\bibfnamefont {S.}~\bibnamefont {Bland}}, \bibinfo
  {author} {\bibfnamefont {G.}~\bibnamefont {Burdiak}}, \bibinfo {author}
  {\bibfnamefont {J.}~\bibnamefont {Chittenden}}, \bibinfo {author}
  {\bibfnamefont {A.}~\bibnamefont {Colaitis}}, \bibinfo {author}
  {\bibfnamefont {P.}~\bibnamefont {De~Grouchy}}, \bibinfo {author}
  {\bibfnamefont {H.}~\bibnamefont {Doyle}}, \bibinfo {author} {\bibfnamefont
  {G.}~\bibnamefont {Hall}},  \emph {et~al.},\ }\href@noop {} {\bibfield
  {journal} {\bibinfo  {journal} {Physical review letters}\ }\textbf {\bibinfo
  {volume} {108}},\ \bibinfo {pages} {145002} (\bibinfo {year}
  {2012})}\BibitemShut {NoStop}%
\bibitem [{\citenamefont {Swadling}\ \emph {et~al.}(2014)\citenamefont
  {Swadling}, \citenamefont {Lebedev}, \citenamefont {Hall}, \citenamefont
  {Patankar}, \citenamefont {Stewart}, \citenamefont {Smith}, \citenamefont
  {Harvey-Thompson}, \citenamefont {Burdiak}, \citenamefont {de~Grouchy},
  \citenamefont {Skidmore} \emph {et~al.}}]{swadling2014diagnosing}%
  \BibitemOpen
  \bibfield  {author} {\bibinfo {author} {\bibfnamefont {G.}~\bibnamefont
  {Swadling}}, \bibinfo {author} {\bibfnamefont {S.}~\bibnamefont {Lebedev}},
  \bibinfo {author} {\bibfnamefont {G.}~\bibnamefont {Hall}}, \bibinfo {author}
  {\bibfnamefont {S.}~\bibnamefont {Patankar}}, \bibinfo {author}
  {\bibfnamefont {N.}~\bibnamefont {Stewart}}, \bibinfo {author} {\bibfnamefont
  {R.}~\bibnamefont {Smith}}, \bibinfo {author} {\bibfnamefont
  {A.}~\bibnamefont {Harvey-Thompson}}, \bibinfo {author} {\bibfnamefont
  {G.}~\bibnamefont {Burdiak}}, \bibinfo {author} {\bibfnamefont
  {P.}~\bibnamefont {de~Grouchy}}, \bibinfo {author} {\bibfnamefont
  {J.}~\bibnamefont {Skidmore}},  \emph {et~al.},\ }\href@noop {} {\bibfield
  {journal} {\bibinfo  {journal} {Review of Scientific Instruments}\ }\textbf
  {\bibinfo {volume} {85}},\ \bibinfo {pages} {11E502} (\bibinfo {year}
  {2014})}\BibitemShut {NoStop}%
\bibitem [{\citenamefont {Ross}\ \emph {et~al.}(2011)\citenamefont {Ross},
  \citenamefont {Divol}, \citenamefont {Sorce}, \citenamefont {Froula},\ and\
  \citenamefont {Glenzer}}]{ross2011ultraviolet}%
  \BibitemOpen
  \bibfield  {author} {\bibinfo {author} {\bibfnamefont {J.}~\bibnamefont
  {Ross}}, \bibinfo {author} {\bibfnamefont {L.}~\bibnamefont {Divol}},
  \bibinfo {author} {\bibfnamefont {C.}~\bibnamefont {Sorce}}, \bibinfo
  {author} {\bibfnamefont {D.}~\bibnamefont {Froula}}, \ and\ \bibinfo {author}
  {\bibfnamefont {S.}~\bibnamefont {Glenzer}},\ }\href@noop {} {\bibfield
  {journal} {\bibinfo  {journal} {Journal of Instrumentation}\ }\textbf
  {\bibinfo {volume} {6}},\ \bibinfo {pages} {P08004} (\bibinfo {year}
  {2011})}\BibitemShut {NoStop}%
\bibitem [{\citenamefont {Pollock}\ \emph {et~al.}(2012)\citenamefont
  {Pollock}, \citenamefont {Meinecke}, \citenamefont {Kuschel}, \citenamefont
  {Ross}, \citenamefont {Shaw}, \citenamefont {Stoafer}, \citenamefont {Divol},
  \citenamefont {Tynan},\ and\ \citenamefont
  {Glenzer}}]{pollock2012simultaneous}%
  \BibitemOpen
  \bibfield  {author} {\bibinfo {author} {\bibfnamefont {B.}~\bibnamefont
  {Pollock}}, \bibinfo {author} {\bibfnamefont {J.}~\bibnamefont {Meinecke}},
  \bibinfo {author} {\bibfnamefont {S.}~\bibnamefont {Kuschel}}, \bibinfo
  {author} {\bibfnamefont {J.}~\bibnamefont {Ross}}, \bibinfo {author}
  {\bibfnamefont {J.}~\bibnamefont {Shaw}}, \bibinfo {author} {\bibfnamefont
  {C.}~\bibnamefont {Stoafer}}, \bibinfo {author} {\bibfnamefont
  {L.}~\bibnamefont {Divol}}, \bibinfo {author} {\bibfnamefont
  {G.}~\bibnamefont {Tynan}}, \ and\ \bibinfo {author} {\bibfnamefont
  {S.}~\bibnamefont {Glenzer}},\ }\href@noop {} {\bibfield  {journal} {\bibinfo
   {journal} {Review of Scientific Instruments}\ }\textbf {\bibinfo {volume}
  {83}},\ \bibinfo {pages} {10E348} (\bibinfo {year} {2012})}\BibitemShut
  {NoStop}%
\bibitem [{\citenamefont {Hall}\ \emph {et~al.}(2014)\citenamefont {Hall},
  \citenamefont {Durmaz}, \citenamefont {Mancini}, \citenamefont {Bailey},
  \citenamefont {Rochau}, \citenamefont {Golovkin},\ and\ \citenamefont
  {MacFarlane}}]{hall2014absorption}%
  \BibitemOpen
  \bibfield  {author} {\bibinfo {author} {\bibfnamefont {I.}~\bibnamefont
  {Hall}}, \bibinfo {author} {\bibfnamefont {T.}~\bibnamefont {Durmaz}},
  \bibinfo {author} {\bibfnamefont {R.}~\bibnamefont {Mancini}}, \bibinfo
  {author} {\bibfnamefont {J.}~\bibnamefont {Bailey}}, \bibinfo {author}
  {\bibfnamefont {G.}~\bibnamefont {Rochau}}, \bibinfo {author} {\bibfnamefont
  {I.}~\bibnamefont {Golovkin}}, \ and\ \bibinfo {author} {\bibfnamefont
  {J.}~\bibnamefont {MacFarlane}},\ }\href@noop {} {\bibfield  {journal}
  {\bibinfo  {journal} {Physics of Plasmas}\ }\textbf {\bibinfo {volume}
  {21}},\ \bibinfo {pages} {031203} (\bibinfo {year} {2014})}\BibitemShut
  {NoStop}%
\bibitem [{\citenamefont {Rochau}\ \emph {et~al.}(2014)\citenamefont {Rochau},
  \citenamefont {Bailey}, \citenamefont {Falcon}, \citenamefont {Loisel},
  \citenamefont {Nagayama}, \citenamefont {Mancini}, \citenamefont {Hall},
  \citenamefont {Winget}, \citenamefont {Montgomery},\ and\ \citenamefont
  {Liedahl}}]{rochau2014zapp}%
  \BibitemOpen
  \bibfield  {author} {\bibinfo {author} {\bibfnamefont {G.~A.}\ \bibnamefont
  {Rochau}}, \bibinfo {author} {\bibfnamefont {J.}~\bibnamefont {Bailey}},
  \bibinfo {author} {\bibfnamefont {R.}~\bibnamefont {Falcon}}, \bibinfo
  {author} {\bibfnamefont {G.}~\bibnamefont {Loisel}}, \bibinfo {author}
  {\bibfnamefont {T.}~\bibnamefont {Nagayama}}, \bibinfo {author}
  {\bibfnamefont {R.}~\bibnamefont {Mancini}}, \bibinfo {author} {\bibfnamefont
  {I.}~\bibnamefont {Hall}}, \bibinfo {author} {\bibfnamefont {D.}~\bibnamefont
  {Winget}}, \bibinfo {author} {\bibfnamefont {M.}~\bibnamefont {Montgomery}},
  \ and\ \bibinfo {author} {\bibfnamefont {D.}~\bibnamefont {Liedahl}},\
  }\href@noop {} {\bibfield  {journal} {\bibinfo  {journal} {Physics of Plasmas
  (1994-present)}\ }\textbf {\bibinfo {volume} {21}},\ \bibinfo {pages}
  {056308} (\bibinfo {year} {2014})}\BibitemShut {NoStop}%
\bibitem [{\citenamefont {Mancini}\ and\ \citenamefont
  {et~al.}(2018)}]{mancini2018}%
  \BibitemOpen
  \bibfield  {author} {\bibinfo {author} {\bibfnamefont {R.}~\bibnamefont
  {Mancini}}\ and\ \bibinfo {author} {\bibnamefont {et~al.}},\ }\href@noop {}
  {\bibfield  {journal} {\bibinfo  {journal} {Submitted for publication}\ }
  (\bibinfo {year} {2018})}\BibitemShut {NoStop}%
\bibitem [{\citenamefont {Froula}\ \emph {et~al.}(2006)\citenamefont {Froula},
  \citenamefont {Ross}, \citenamefont {Divol},\ and\ \citenamefont
  {Glenzer}}]{froula2006thomson}%
  \BibitemOpen
  \bibfield  {author} {\bibinfo {author} {\bibfnamefont {D.}~\bibnamefont
  {Froula}}, \bibinfo {author} {\bibfnamefont {J.}~\bibnamefont {Ross}},
  \bibinfo {author} {\bibfnamefont {L.}~\bibnamefont {Divol}}, \ and\ \bibinfo
  {author} {\bibfnamefont {S.}~\bibnamefont {Glenzer}},\ }\href@noop {}
  {\bibfield  {journal} {\bibinfo  {journal} {Review of scientific
  instruments}\ }\textbf {\bibinfo {volume} {77}},\ \bibinfo {pages} {10E522}
  (\bibinfo {year} {2006})}\BibitemShut {NoStop}%
\bibitem [{\citenamefont {Sheffield}\ \emph {et~al.}(2010)\citenamefont
  {Sheffield}, \citenamefont {Froula}, \citenamefont {Glenzer},\ and\
  \citenamefont {Luhmann~Jr}}]{sheffield2010plasma}%
  \BibitemOpen
  \bibfield  {author} {\bibinfo {author} {\bibfnamefont {J.}~\bibnamefont
  {Sheffield}}, \bibinfo {author} {\bibfnamefont {D.}~\bibnamefont {Froula}},
  \bibinfo {author} {\bibfnamefont {S.~H.}\ \bibnamefont {Glenzer}}, \ and\
  \bibinfo {author} {\bibfnamefont {N.~C.}\ \bibnamefont {Luhmann~Jr}},\
  }\href@noop {} {\emph {\bibinfo {title} {Plasma scattering of electromagnetic
  radiation: theory and measurement techniques}}}\ (\bibinfo  {publisher}
  {Academic press},\ \bibinfo {year} {2010})\BibitemShut {NoStop}%
\bibitem [{\citenamefont {Jones}\ \emph {et~al.}(01  )\citenamefont {Jones},
  \citenamefont {Oliphant}, \citenamefont {Peterson} \emph {et~al.}}]{scipy}%
  \BibitemOpen
  \bibfield  {author} {\bibinfo {author} {\bibfnamefont {E.}~\bibnamefont
  {Jones}}, \bibinfo {author} {\bibfnamefont {T.}~\bibnamefont {Oliphant}},
  \bibinfo {author} {\bibfnamefont {P.}~\bibnamefont {Peterson}},  \emph
  {et~al.},\ }\href {http://www.scipy.org/} {\enquote {\bibinfo {title}
  {{SciPy}: Open source scientific tools for {Python}},}\ } (\bibinfo {year}
  {2001--})\BibitemShut {NoStop}%
\bibitem [{\citenamefont {{The PlasmaPy Community}}(17  )}]{plasmapy}%
  \BibitemOpen
  \bibfield  {author} {\bibinfo {author} {\bibnamefont {{The PlasmaPy
  Community}}},\ }\href {https://github.com/PlasmaPy/PlasmaPy} {\enquote
  {\bibinfo {title} {{PlasmaPy}: An open source community developed {Python}
  package for plasma physics},}\ } (\bibinfo {year} {2017--})\BibitemShut
  {NoStop}%
\bibitem [{\citenamefont {Ross}\ \emph {et~al.}(2010)\citenamefont {Ross},
  \citenamefont {Glenzer}, \citenamefont {Palastro}, \citenamefont {Pollock},
  \citenamefont {Price}, \citenamefont {Tynan},\ and\ \citenamefont
  {Froula}}]{ross2010thomson}%
  \BibitemOpen
  \bibfield  {author} {\bibinfo {author} {\bibfnamefont {J.}~\bibnamefont
  {Ross}}, \bibinfo {author} {\bibfnamefont {S.}~\bibnamefont {Glenzer}},
  \bibinfo {author} {\bibfnamefont {J.}~\bibnamefont {Palastro}}, \bibinfo
  {author} {\bibfnamefont {B.}~\bibnamefont {Pollock}}, \bibinfo {author}
  {\bibfnamefont {D.}~\bibnamefont {Price}}, \bibinfo {author} {\bibfnamefont
  {G.}~\bibnamefont {Tynan}}, \ and\ \bibinfo {author} {\bibfnamefont
  {D.}~\bibnamefont {Froula}},\ }\href@noop {} {\bibfield  {journal} {\bibinfo
  {journal} {Review of Scientific Instruments}\ }\textbf {\bibinfo {volume}
  {81}},\ \bibinfo {pages} {10D523} (\bibinfo {year} {2010})}\BibitemShut
  {NoStop}%
\bibitem [{\citenamefont {Rambo}\ \emph {et~al.}(2016)\citenamefont {Rambo},
  \citenamefont {Schwarz}, \citenamefont {Schollmeier}, \citenamefont
  {Geissel}, \citenamefont {Smith}, \citenamefont {Kimmel}, \citenamefont
  {Speas}, \citenamefont {Shores}, \citenamefont {Armstrong}, \citenamefont
  {Bellum} \emph {et~al.}}]{rambo2016sandia}%
  \BibitemOpen
  \bibfield  {author} {\bibinfo {author} {\bibfnamefont {P.}~\bibnamefont
  {Rambo}}, \bibinfo {author} {\bibfnamefont {J.}~\bibnamefont {Schwarz}},
  \bibinfo {author} {\bibfnamefont {M.}~\bibnamefont {Schollmeier}}, \bibinfo
  {author} {\bibfnamefont {M.}~\bibnamefont {Geissel}}, \bibinfo {author}
  {\bibfnamefont {I.}~\bibnamefont {Smith}}, \bibinfo {author} {\bibfnamefont
  {M.}~\bibnamefont {Kimmel}}, \bibinfo {author} {\bibfnamefont
  {C.}~\bibnamefont {Speas}}, \bibinfo {author} {\bibfnamefont
  {J.}~\bibnamefont {Shores}}, \bibinfo {author} {\bibfnamefont
  {D.}~\bibnamefont {Armstrong}}, \bibinfo {author} {\bibfnamefont
  {J.}~\bibnamefont {Bellum}},  \emph {et~al.},\ }in\ \href@noop {} {\emph
  {\bibinfo {booktitle} {Laser-Induced Damage in Optical Materials 2016}}},\
  Vol.\ \bibinfo {volume} {10014}\ (\bibinfo {organization} {International
  Society for Optics and Photonics},\ \bibinfo {year} {2016})\ p.\ \bibinfo
  {pages} {100140Z}\BibitemShut {NoStop}%
\bibitem [{\citenamefont {Wood}(2003)}]{wood2003laser}%
  \BibitemOpen
  \bibfield  {author} {\bibinfo {author} {\bibfnamefont {R.~M.}\ \bibnamefont
  {Wood}},\ }\href@noop {} {\emph {\bibinfo {title} {Laser-induced damage of
  optical materials}}}\ (\bibinfo  {publisher} {CRC Press},\ \bibinfo {year}
  {2003})\BibitemShut {NoStop}%
\bibitem [{\citenamefont {Falcon}\ \emph {et~al.}(2013)\citenamefont {Falcon},
  \citenamefont {Rochau}, \citenamefont {Bailey}, \citenamefont {Ellis},
  \citenamefont {Carlson}, \citenamefont {Gomez}, \citenamefont {Montgomery},
  \citenamefont {Winget}, \citenamefont {Chen}, \citenamefont {Gomez} \emph
  {et~al.}}]{falcon2013experimental}%
  \BibitemOpen
  \bibfield  {author} {\bibinfo {author} {\bibfnamefont {R.~E.}\ \bibnamefont
  {Falcon}}, \bibinfo {author} {\bibfnamefont {G.}~\bibnamefont {Rochau}},
  \bibinfo {author} {\bibfnamefont {J.}~\bibnamefont {Bailey}}, \bibinfo
  {author} {\bibfnamefont {J.}~\bibnamefont {Ellis}}, \bibinfo {author}
  {\bibfnamefont {A.}~\bibnamefont {Carlson}}, \bibinfo {author} {\bibfnamefont
  {T.}~\bibnamefont {Gomez}}, \bibinfo {author} {\bibfnamefont
  {M.}~\bibnamefont {Montgomery}}, \bibinfo {author} {\bibfnamefont
  {D.}~\bibnamefont {Winget}}, \bibinfo {author} {\bibfnamefont
  {E.}~\bibnamefont {Chen}}, \bibinfo {author} {\bibfnamefont {M.}~\bibnamefont
  {Gomez}},  \emph {et~al.},\ }\href@noop {} {\bibfield  {journal} {\bibinfo
  {journal} {High Energy Density Physics}\ }\textbf {\bibinfo {volume} {9}},\
  \bibinfo {pages} {82} (\bibinfo {year} {2013})}\BibitemShut {NoStop}%
\bibitem [{\citenamefont {Falcon}\ \emph {et~al.}(2015)\citenamefont {Falcon},
  \citenamefont {Rochau}, \citenamefont {Bailey}, \citenamefont {Gomez},
  \citenamefont {Montgomery}, \citenamefont {Winget},\ and\ \citenamefont
  {Nagayama}}]{falcon2015laboratory}%
  \BibitemOpen
  \bibfield  {author} {\bibinfo {author} {\bibfnamefont {R.~E.}\ \bibnamefont
  {Falcon}}, \bibinfo {author} {\bibfnamefont {G.~A.}\ \bibnamefont {Rochau}},
  \bibinfo {author} {\bibfnamefont {J.~E.}\ \bibnamefont {Bailey}}, \bibinfo
  {author} {\bibfnamefont {T.~A.}\ \bibnamefont {Gomez}}, \bibinfo {author}
  {\bibfnamefont {M.~H.}\ \bibnamefont {Montgomery}}, \bibinfo {author}
  {\bibfnamefont {D.~E.}\ \bibnamefont {Winget}}, \ and\ \bibinfo {author}
  {\bibfnamefont {T.}~\bibnamefont {Nagayama}},\ }\href@noop {} {\bibfield
  {journal} {\bibinfo  {journal} {The Astrophysical Journal}\ }\textbf
  {\bibinfo {volume} {806}},\ \bibinfo {pages} {214} (\bibinfo {year}
  {2015})}\BibitemShut {NoStop}%
\bibitem [{\citenamefont {Kramida}\ \emph {et~al.}(2018)\citenamefont
  {Kramida}, \citenamefont {Ralchenko}, \citenamefont {Reader},\ and\
  \citenamefont {{NIST ASD Team}}}]{kramida2018nist}%
  \BibitemOpen
  \bibfield  {author} {\bibinfo {author} {\bibfnamefont {A.}~\bibnamefont
  {Kramida}}, \bibinfo {author} {\bibfnamefont {Y.}~\bibnamefont {Ralchenko}},
  \bibinfo {author} {\bibfnamefont {J.}~\bibnamefont {Reader}}, \ and\ \bibinfo
  {author} {\bibnamefont {{NIST ASD Team}}},\ }\href
  {https://physics.nist.gov/asd} {\enquote {\bibinfo {title} {{NIST} atomic
  spectra database (ver. 5.5.2)},}\ } (\bibinfo {year} {2018})\BibitemShut
  {NoStop}%
\bibitem [{\citenamefont {MacFarlane}\ \emph {et~al.}(2006)\citenamefont
  {MacFarlane}, \citenamefont {Golovkin},\ and\ \citenamefont
  {Woodruff}}]{macfarlane2006helios}%
  \BibitemOpen
  \bibfield  {author} {\bibinfo {author} {\bibfnamefont {J.}~\bibnamefont
  {MacFarlane}}, \bibinfo {author} {\bibfnamefont {I.}~\bibnamefont
  {Golovkin}}, \ and\ \bibinfo {author} {\bibfnamefont {P.}~\bibnamefont
  {Woodruff}},\ }\href@noop {} {\bibfield  {journal} {\bibinfo  {journal}
  {Journal of Quantitative Spectroscopy and Radiative Transfer}\ }\textbf
  {\bibinfo {volume} {99}},\ \bibinfo {pages} {381} (\bibinfo {year}
  {2006})}\BibitemShut {NoStop}%
\bibitem [{\citenamefont {Swadling}\ \emph {et~al.}(2017)\citenamefont
  {Swadling}, \citenamefont {Ross}, \citenamefont {Manha}, \citenamefont
  {Galbraith}, \citenamefont {Datte}, \citenamefont {Sorce}, \citenamefont
  {Katz}, \citenamefont {Froula}, \citenamefont {Widmann}, \citenamefont
  {Jones} \emph {et~al.}}]{swadling2017initial}%
  \BibitemOpen
  \bibfield  {author} {\bibinfo {author} {\bibfnamefont {G.}~\bibnamefont
  {Swadling}}, \bibinfo {author} {\bibfnamefont {J.}~\bibnamefont {Ross}},
  \bibinfo {author} {\bibfnamefont {D.}~\bibnamefont {Manha}}, \bibinfo
  {author} {\bibfnamefont {J.}~\bibnamefont {Galbraith}}, \bibinfo {author}
  {\bibfnamefont {P.}~\bibnamefont {Datte}}, \bibinfo {author} {\bibfnamefont
  {C.}~\bibnamefont {Sorce}}, \bibinfo {author} {\bibfnamefont
  {J.}~\bibnamefont {Katz}}, \bibinfo {author} {\bibfnamefont {D.}~\bibnamefont
  {Froula}}, \bibinfo {author} {\bibfnamefont {K.}~\bibnamefont {Widmann}},
  \bibinfo {author} {\bibfnamefont {O.}~\bibnamefont {Jones}},  \emph
  {et~al.},\ }\href@noop {} {\bibfield  {journal} {\bibinfo  {journal} {Physics
  of Plasmas}\ }\textbf {\bibinfo {volume} {24}},\ \bibinfo {pages} {032705}
  (\bibinfo {year} {2017})}\BibitemShut {NoStop}%
\end{thebibliography}%
	
%
%
%
	
\end{document}